\documentclass[twocolumn,showpacs,preprintnumbers,nofootinbib
amsmath,amssymb,aps,prl]{revtex4}
\usepackage{dcolumn}
\usepackage{amsthm,amscd,mathrsfs}
\usepackage{amsfonts,bm,latexsym}
\def\be{\begin{equation}}
\def\ee{\end{equation}}
\def\bea{\begin{eqnarray}}
\def\eea{\end{eqnarray}}
\def\nn{\nonumber}
\def\p{\partial}

\begin{document}
\preprint{arXiv:0709.1749v4 [hep-th]}

\title{General Non-extremal Rotating Charged G\"{o}del Black Holes in Minimal
Five-Dimensional Gauged Supergravity}

\author{Shuang-Qing Wu}\email{sqwu@phy.ccnu.edu.cn}

\affiliation{\centerline{College of Physical Science and Technology,
Central China Normal University, Wuhan, Hubei 430079, People's Republic of China}}
\date{Submitted 11 September 2007; Revised 7 December 2007}

\begin{abstract}
We present the general exact solutions for non-extremal rotating charged black holes in the G\"{o}del
universe of five-dimensional minimal supergravity theory. They are uniquely characterized by four
non-trivial parameters, namely the mass $m$, the charge $q$, the Kerr equal rotation parameter $a$,
and the G\"{o}del parameter $j$. We calculate the conserved energy, angular momenta and charge for
the solutions and show that they completely satisfy the first law of black hole thermodynamics. We
also study the symmetry and separability of the Hamilton-Jacobi and the massive Klein-Gordon equations
in these Einstein-Maxwell-Chern-Simons-G\"{o}del black hole backgrounds.
\end{abstract}

\pacs{04.70.Dy, 04.20.Jb, 04.65.+e, 11.25.Wx}

\maketitle

\textit{Introduction}.---%
A realistic black hole must localize inside the cosmological background, and a natural consideration
for it is our universe (with a possible global rotation). A popular phenomenological theory to describe
the universe is given by the standard Friedman-Robertson-Walker (FRW) metric, which represents a rather
idealized model of an isotropic homogeneous world filled with perfect fluid. But the standard FRW model
is too ideal to describe the global rotation of the universe, since the rotation is a universal phenomenon:
all compact objects in the universe rotate.

An exact solution for the rotating universe in four dimensions was found by G\"{o}del \cite{Godel}.
The G\"{o}del universe is an exact solution of Einstein's equation in the presence of a cosmological
constant and homogeneous pressureless matter. This space-time solution exhibits several peculiar
features including in particular the presence of closed timelike curves (CTCs) through every point.

Recently, the solutions representing the generalization of the G\"{o}del universe to $D = 5$ dimensions,
especially in the context of five-dimensional minimal supergravity, have attracted a lot of attention
\cite{GGHPR,GH,CARH,KNG,BDGO,BC,KV,KM,KA,CGLP,BGHV,TO}. A maximally supersymmetric G\"{o}del-type
universe \cite{GGHPR} exhibiting most of the peculiar features of the four-dimensional cousin was shown
to be an exact solution of minimal supergravity in $D = 5$ dimensions. These solutions are related by
T-duality to pp-waves, when uplifted to 10 dimensions \cite{BGHV}. The G\"{o}del-type universes are
important due to the possibility of quantizing strings in these backgrounds and their relation to the
corresponding limits of super-Yang Mills theories. On the gravitational side, the pp-waves (U-)dual
to the G\"{o}del universes arise as the Penrose limits of near-horizon geometries. Quite recently,
an exact solution for a stationary Kerr black hole immersed in the rotating G\"{o}del universe has
been obtained by Gimon and Hashimoto \cite{GH} within the five-dimensional minimal supergravity,
and its various properties have been investigated recently in \cite{BC,KV,KM,KA,CGLP}. A procedure
was proposed \cite{TO,KLa} to embed the supersymmetric black ring solutions into the G\"{o}del universe,
but no explicit solution was presented.

On the other hand, it is very difficult to find an exact rotating charged solution in higher dimensions.
Recently, there has been renewed interest \cite{BMPV,GMT,KS,CLP,MR,KLb,Ross,CCLP} in finding rotating
charged solutions of five-dimensional gravity and supergravity. The BMPV black hole \cite{BMPV,GMT}
is the only asymptotically flat, half supersymmetric, rotating charged black hole with a regular, finite
size horizon and finite entropy. Its existence is made possible due to the particular Chern-Simons coupling
of minimal $D = 5$ supergravity and the fact that a self-duality condition is allowed to impose on the
exterior derivative of the rotation one-form in $D = 5$ dimensions. Later on, further generalizations
have been made in \cite{KS,CLP,MR,KLb,Ross,CCLP} to include a nonzero cosmological constant.

As far as we are aware, until recently, an exact solution for a rotating charged black hole localized
inside the G\"{o}del universe was not known. A charged extremal black hole with finite horizon area in
a G\"{o}del universe was identified in \cite{GGHPR,CARH}, where both the pure G\"{o}del-type universe
and the BMPV black hole had also been discussed, but they were presented in two separate solutions, not
in a single form. The Kerr-Newman-G\"{o}del black hole \cite{KNG} and its three-charge's generalization
\cite{BDGO} are not exact solutions of $D = 5$ minimal supergravity, but in the extremal limit they
indeed yields a remarkable superimposition of the BMPV black hole and the pure G\"{o}del-type universe
in one solution, namely the so-called BMPV-G\"{o}del black hole. Therefore, it is an important outstanding
problem to find space-times describing a non-extremal rotating charged black hole immersed in the
G\"{o}del universe.

In this Letter, we construct an exact solution describing the non-extremal rotating charged black
hole localized inside the G\"{o}del universe, and describe its various basic properties. A remarkable
feature of the solution is that in the $m = q$ case it superimposes the pure G\"{o}del-type universe
and the BMPV black hole in one solution, similar to that in \cite{KNG}. Our solution, however, is a
faith, charged generalization of the Kerr-G\"{o}del black holes found by Gimon and Hashimotoin \cite{GH}.
We shall refer to our solution as the Einstein-Maxwell-Chern-Simons-G\"{o}del (EMCS-G\"{o}del) black
hole, in order to distinguish it from the one derived in \cite{KNG}. Just like \cite{GH}, we also do
not require our solutions to preserve any supersymmetry \cite{GGHPR}. Another important property is
that these EMCS-G\"{o}del black holes still obey the first law (integral and differential Bekenstein-Smarr
formulae) of black hole thermodynamics provided the G\"{o}del parameter $j$ is viewed as a thermodynamical
variable. In addition, the Hamilton-Jacobi and Klein-Gordon equations are separable in these backgrounds
and the space-time admits a reducible Killing tensor.

\textit{Metric ansatz}.---%
The bosonic part of the minimal supergravity theory in $D = 5$ dimensions consists of the metric
and a one-form gauge field, which are governed by the EMCS equations of motion
\bea
 && R_{\mu\nu} -\frac{1}{2}g_{\mu\nu}R = 2\Big(F_{\mu\alpha}F_{\nu}^{~\alpha}
  -\frac{1}{4}g_{\mu\nu}F_{\alpha\beta}F^{\alpha\beta}\Big) \, , \label{Einstein} \qquad \\
 && D_{\nu}\Big(F^{\mu\nu} +\frac{\lambda}{\sqrt{3}\sqrt{-g}}
  \epsilon^{\mu\nu\alpha\beta\gamma} A_{\alpha}F_{\beta\gamma}\Big) = 0 \, . \label{MCS}
\eea
where $\epsilon^{\mu\nu\alpha\beta\gamma}$ is the Levi-Civita tensor density, and $\lambda = 1$ is
the Chern-Simons coupling constant.

To seek an exact solution of the supergravity equations of motion, it is of particular importance
to start from a good ansatz for the metric and the gauge potential. A suitable ansatz to our aim
assumes the following form
\bea
 ds^2 &=& -f(r)dt^2 -2g(r)\sigma_3dt +h(r)\sigma_3^2 \nn \\
 && +\frac{dr^2}{V(r)} +\frac{r^2}{4}d\Omega_3^2 \, , \label{mansatz} \\
 A &=& B(r)dt +C(r)\sigma_3 \, , \label{gansatz}
\eea
where the unit $3$-sphere $d\Omega_3^2$ and the left invariant form $\sigma_3$ are specified by
\bea
 d\Omega_3^2 
 = d\theta^2 +\sin^2\theta d\psi^2 +\sigma_3^2 \, , \quad
 \sigma_3 = d\phi +\cos\theta d\psi \, .
\eea

To arrive at the above ansatz for charged generalization of the Kerr-G\"{o}del black hole, we have
compared two already-known solutions (a uncharged metric and a charged solution) for the rotating
black holes in minimal $D = 5$ supergravity. The first uncharged solution \cite{GH} represents the
$D = 5$ Kerr black holes embedded in the rotating G\"{o}del universe, its explicit form is given
by Eq. (\ref{ckgbh}) below by setting the charge $q = 0$. In this metric, the parameter $j$ defines
the scale of the G\"{o}del background and is responsible for the rotation of the universe. The parameter
$a$ is related to the rotation of the black hole. When $m = a = 0$, the metric reduces to that of the
pure G\"{o}del-type universe \cite{GGHPR}. The $D = 5$ Kerr black hole with equal rotation parameters
is recovered when $j = 0$.

The second one comes from the charged supergravity solutions in the vacuum backgrounds. The solution for
a rotating charged Kerr black hole in five dimensions is again given by Eq. (\ref{ckgbh}) by taking $j =
0$. This solution is the five-dimensional Kerr-Newman black hole with equal-magnitude angular momenta,
satisfying the EMCS equations (\ref{Einstein}) and (\ref{MCS}). It is a correct generalization of the
four-dimensional Kerr-Newman solution to five dimensions. When $a = 0$, it reduces to the $D = 5$
Reissner-Nordstr\"{o}m black hole. The $m = q$ case reproduces the supersymmetric BMPV black hole
\cite{BMPV,GMT}. This charged solution is also related to the previously known charged solutions
\cite{KS,CLP,MR,KLb,Ross,CCLP} in the case without a cosmological constant. In particular, it corresponds
to the solution in \cite{CLP} by taking $\beta = \lambda = 0$ and to that in \cite{MR,KLb,Ross} by $m =
p -q$ and $\lambda = 0$. It can also be reduced from the general solution in \cite{CCLP} (with $g = 0$)
by setting two angular momenta equal and by redefining the coordinates. The $m = q$ case coincides with
the solution in \cite{KS} when $g = 0$.

Clearly, our metric ansatz keeps five of the nine isometries of the G\"{o}del universe, generated by
$\p_t$, and by four generators of the SU$(2)$ $\times$ U$(1)$ subgroup of the SO$(4)$ isometry group
acting on $S^3$ \cite{GH}. Our ansatz is also inspired from symmetry and separability of various field
equations.

\textit{The solutions
and basic properties}.---%
We now try to seek an exact charged solution representing the EMCS-G\"{o}del black hole. There are six
unknown functions needed to be specified. But among them, a constraint
\be
 V(r) = 4\frac{g(r)^2 +h(r)f(r)}{r^2} +f(r) \, ,
\ee
will reduce the actual number to five.

Substituting the ansatz (\ref{mansatz}) and (\ref{gansatz}) into Eqs. (\ref{Einstein}) and (\ref{MCS})
will result in a rather complicated set of equations. We are guided by the fact that our solution
should reduce to the Kerr-G\"{o}del black hole \cite{GH} when the charge parameter vanishes, and
to the Kerr-Newman solution when the G\"{o}del parameter is equal to zero. With this in mind, we
assume that each unknown function is a polynomial of the radial coordinate, with its coefficients
to be determined. Using the GRTensor II program, it is not difficult to check that the following
choice (with $\lambda = 1$)
\bea
 f(r) &=& 1 -\frac{2m}{r^2} +\frac{q^2}{r^4} \, , \nn \\
 g(r) &=& jr^2 +3jq +\frac{(2m -q)a}{2r^2} -\frac{q^2a}{2r^4} \, , \nn \\
 h(r) &=& -j^2r^2(r^2 +2m +6q) +3jqa \nn \\
 && +\frac{(m -q)a^2}{2r^2} -\frac{q^2a^2}{4r^4} \, , \nn \\
 V(r) &=& 1 -\frac{2m}{r^2} +\frac{8j(m +q)\big[a +2j(m +2q)\big]}{r^2} \nn \\
 &+& \frac{2(m -q)a^2 +q^2\big[1 -16ja -8j^2(m +3q)\big]}{r^4} \, , \nn \quad \\
 B(r) &=& \frac{\sqrt{3}q}{2r^2} \, , \quad
 C(r) = \frac{\sqrt{3}}{2}\Big(jr^2 +2jq -\frac{qa}{2r^2}\Big) \, , \label{ckgbh}
\eea
indeed solves the EMCS equations (\ref{Einstein}) and (\ref{MCS}).

The solution presented above is the `Left' form. Its corresponding `Right' solution can be generated
by the following dual transformations ($\lambda = 1$)
\bea
\psi \leftrightarrow \phi \, , \quad~ q \to -q \, , \quad j \to -j \, ,
\quad~ \lambda \to -\lambda \, .
\eea

In the remaining analysis, we shall focus on the `Left' solution only. The solution is, in general,
non-extremal. It is uniquely characterized by four non-trivial parameters, namely the mass $m$, the
charge $q$, the equal Kerr rotation parameter $a$, and the G\"{o}del parameter $j$. When $j = 0$,
the metric reduces to the $D = 5$ Kerr-Newman solution. In the case when $q = 0$, we recover the
Kerr-G\"{o}del black hole \cite{GH}. When the parameter $a$ is set to zero, the solution represents
a non-extremal $D =5$ Reissner-Nordstr\"{o}m-G\"{o}del black hole. A supersymmetric G\"{o}del black
hole appears when $m = -q$, $a = -2jq$, corresponding to the charged extremal G\"{o}del black hole
previously identified in \cite{GGHPR,CARH}. The $m = q$ case is in particular interesting, it is the
BMPV-G\"{o}del black hole, a superimposition of two remarkable solutions in minimal $D = 5$ supergravity,
that is, the pure G\"{o}del-type universe and the BMPV black hole. This charged Kerr-G\"{o}del solution
with equal rotation parameters is one of the main results in this Letter. In what follows, we shall
study its various basic properties.

The space-time has a curvature singularity at $r = 0$, where both the Ricci scalar and the gauge
field strength
\bea
 &R& = \frac{1}{3}F_{\mu\nu}F^{\mu\nu}
 = 16j^2\Big(1 -\frac{m -3q}{r^2} +\frac{6q^2}{r^4}\Big) \nn \\
 &&~~ -\frac{2q^2\big[1 +8ja -8j^2(m +3q)\big]}{r^6} +\frac{4q^2a^2}{r^8} \, , \qquad
\eea
diverge there.

A salient feature of the solution (\ref{ckgbh}) is that the space-time has an event horizon at $r_+$
and an inner horizon at $r_-$, which are determined by $V(r) = 0$, namely, the locations of black hole
horizons are
\bea
 r_{\pm}^2 &=& m -4j(m +q)a -8j^2(m +q)(m +2q) \pm \sqrt{\delta} \, , \nn \\
 \delta &=& \big[m -q -8j^2(m +q)^2\big]\big[m +q -2a^2 \nn \\
 &&\quad -8j(m +2q)a -8j^2(m +2q)^2 \big] \, .
\eea

The metric is well behaved at the horizons but the gauge field becomes singular there \cite{CGLP}.
Clearly $\delta > 0$ is the condition for the horizon to be well defined. The horizons will degenerate
when $\delta = 0$, corresponding to the following different possibilities for $j$ happens,
\bea
 j_{\pm} = \pm \frac{\sqrt{2(m -q)}}{4(m +q)} \, ; \quad {\rm or} \quad
 \frac{-2a \pm \sqrt{2(m +q)}}{4(m +2q)} \, .
\eea
In these cases, the black hole becomes extremal. On the other hand, a naked singularity will appear
when $\delta < 0$.

Two ergospheres appear at $r_{erg}^2 = m \pm \sqrt{m^2 -q^2}$ where the function $f(r)$ vanishes.
They will coincide with each other at $m = \pm q$.

Just as their uncharged counterparts, the EMCS-G\"{o}del solutions can have CTCs. The surface
at fixed $r$ where the metric component $g_{\phi\phi} = h(r) +r^2/4$ vanishes is called as the
``velocity of light surface'' (VLS) or the CTC horizon. The location of CTC horizon $r = r_{\rm
VLS}$ is determined by $h(r) +r^2/4 = 0$, namely
\bea
 && r^6\big[1 -4j^2(r^2 +2m +6q)\big] +12jqar^4 \qquad \nn \\
 &&\qquad\qquad +2(m -q)a^2r^2 -q^2a^2 = 0 \, . \label{ctc}
\eea
If $r > r_{\rm VLS}$ (when $g_{\phi\phi} < 0$), then $\p_{\phi}$ will be timelike, indicating
the presence of CTCs since $\phi$ is periodic. There will be no CTCs for $r < r_{\rm VLS}$ (when
$g_{\phi\phi} > 0$). It should be emphasized that since the G\"{o}del space-time is homogeneous,
there is a CTC going through every point in space-time, i.e. the time machine.

The roots of the CTC horizon equation (\ref{ctc}) are, in general, very complicated. However, two
special cases are relatively simple. In the case of a Reissner-Nordstr\"{o}m-G\"{o}del black hole
(when $a = 0$), the velocity of light surface is situated at $r_{\rm VLS}^2 = 1/(4j^2) -2m -6q$.
By contrast, in the case where $m = q = 1/(32j^2)$, the CTC horizons are located at $r_{\rm VLS}^2
= (\sqrt{2} \pm 1)a^{1/2}/(8j^{3/2})$.

It is recognized that for the existence of regular EMCS-G\"{o}del black holes, the four parameters
must lie in appropriate ranges so that naked singularities and CTCs are avoided. In other words,
they must simultaneously satisfy $\delta > 0$ and $g_{\phi\phi} = h(r) +r^2/4 > 0$. Compared with
the uncharged version \cite{GH}, the parameter space for the charged Kerr-G\"{o}del black hole is
much richer than that of a neutral Kerr-G\"{o}del black hole. Therefore it deserves a further analysis
in detail, as did in \cite{KM}.

\textit{Thermodynamics of
EMCS-G\"{o}del black holes}.---%
For a regular rotating EMCS-G\"{o}del black hole, the horizon topology is a squashed $3$-sphere. We
now investigate its thermodynamical properties. The Bekenstein-Hawking entropy of the black hole is
\be
 S = \frac{1}{4}\mathcal{A} = \frac{1}{2}\pi^2r_+^2\sqrt{4h(r_+) +r_+^2} \, ,
\ee
while the Hawking temperature $T = \kappa/(2\pi)$ is given via the surface gravity
\be
 \kappa = \frac{r_+V'(r_+)}{2\sqrt{4h(r_+) +r_+^2}}
  = \frac{r_+^2 -r_-^2}{r_+^2\sqrt{4h(r_+) +r_+^2}} \, .
\ee
The latter can be obtained by a standard Wick-rotation approach or computed via the standard formula
$\kappa^2 = -\frac{1}{2}l_{\mu;\nu}l^{\mu;\nu}|_{r=r_+}$, where the Killing vector $l = \p_t +\Omega
\p_{\phi}$ is normal to and becomes null at the horizon.

The angular velocity $\Omega$ and the electrostatic potential $\Phi$ at the horizon are given by
\bea
 && \Omega = \Omega_{\phi} = g(r_+)/\big[h(r_+) +r_+^2/4\big] \, ,
 \quad~~ \Omega_{\psi} = 0 \, , \qquad\quad \\
 && \Phi = l^{\mu}A_{\mu}\big|_{r=r_+} = B(r_+) +C(r_+)\Omega \, .
\eea
There is a special choice of parameters if they satisfy
\be
 (m -q)a^2 +4j(m -q)(m +2q)a -4j^2(3m +5q)q^2 = 0 \, ,
\ee
then $V(r_+) = g(r_+) = 0$. Consequently $\Omega$ will vanish at the horizon. This generalizes
the non-trivial result ($a = -4jm$) in the case of a Kerr-G\"{o}del black hole \cite{GH}.

It is remarkable that the conserved energy, angular momenta and charge for the charged Kerr-G\"{o}del
black hole,
\bea
 M &=& \pi\Big[\frac{3}{4}m -j(m +q)a -2j^2(m +q)(4m +5q)\Big] \, , ~~ \nn \\
 J &=& \frac{1}{2}\pi\Big\{a\big[m -\frac{q}{2} -2j(m -q)a -8j^2(m^2 +mq \nn \\
 &&\quad -2q^2)\big] -3jq^2 +8j^2(3m +5q)q^2\Big\} \, , \nn \\
 Q &=& \frac{\sqrt{3}}{2}\pi\big[q -4j(m +q)a -8j^2(m +q)q\big] \, , \nn \\
 W &=& 2\pi\big(m +q\big)\big[a +2j(m +2q)\big] \, ,
\eea
satisfy the first law of black hole thermodynamics
\bea
 d M &=& T dS +\Omega dJ +\Phi dQ +W dj \, , \qquad \\
 \frac{2}{3}M &=& T S +\Omega J +\frac{2}{3}\Phi Q -\frac{1}{3}W j \, .
\eea

To close the integral Bekenstein-Smarr formula, here we have considered the G\"{o}del parameter $j$ as
a thermodynamical variable, and introduced its conjugate generalized force $W$. The conserved charges
for $M$, $J$, and $Q$ were computed at first by integrating the first law via fixing the parameter $j$
as a constant. Once they are determined, then we can allow $j$ to vary and check the first laws to
obtain the expression for $W$. It is also possible to use ($Wj^2$, $1/j$) as a pair of thermodynamical
variables to change the sign of the G\"{o}del work term $Wj$. Compared with the thermodynamical role
\cite{KLb,BTZbhT} played in by the cosmological constant, we can argue that the G\"{o}del parameter
behaves just like a cosmological constant in the sense of thermodynamics.

\textit{Symmetry and separability of
Hamilton-Jacobi and Klein-Gordon equations}.---%
From the inverse metric components and the metric determinant, one can easily infer that not only the
Hamilton-Jacobi equation, but also the massive scalar equation with a minimal electro-magnetic coupling
term, are capable of separation of variables. This is contrary to the statement made in \cite{KA}, where
the non-separability of Klein-Gordon equation is because the authors had adopted a different coordinate
system and expanded the metric in the regime for small $j$. The separability arises from the fact that
the metric ansatz keeps the five isometries of the G\"{o}del universe, generated by $\p_t$ and four
generators of the SU$(2)$ $\times$ U$(1)$ group acting on $S^3$ \cite{GH}. The separability also implies
that the metric admits a reducible Killing tensor \cite{KLb}. Details will be published elsewhere.

\textit{Discussions}.---%
There are many other interesting issues to explore. The computation of the conserved energy, angular
momenta and electric charge of the EMCS-G\"{o}del black holes remains a big challenge, because the
naive application of traditional approaches such as the counter-term method fails. At present, there
is only one viable work \cite{BC} that can do such a job. It remains an open question whether the
conserved charges can be computed by other well-known methods. One would like to map out the full
parameter spaces of the general non-extremal charged Kerr-G\"{o}del solution. One can lift the $D
= 5$ EMCS-G\"{o}del solution to 10 dimensions as a new background for string and M-theory. Ultimately,
it would also be interesting to study the causality problem, chronology protection and holography
\cite{BGHV}. We hope that the explicit solution describing the non-extremal charged Kerr black holes
immersed in the rotating G\"{o}del-type universe will stimulate further insight into these fascinating
issues.

\begin{acknowledgments}
This work is partially supported by the NSFC under Grant No. 10675051.
\end{acknowledgments}


\vskip -8pt

\begin{thebibliography}{99}

\bibitem{Godel}
K. G\"{o}del, Rev. Mod. Phys. \textbf{21}, 447 (1949).

\bibitem{GGHPR}
J.P. Gauntlett, J.B. Gutowski, C.M. Hull, S. Pakis, and H.S. Reall, Class. Quantum Grav. \textbf{20}, 4587 (2003).

\bibitem{CARH}
C.A.R. Herdeiro, Nucl. Phys. \textbf{B665}, 189 (2003).

\bibitem{GH}
E.G. Gimon and A. Hashimoto, Phys. Rev. Lett. \textbf{91}, 021601 (2003).

\bibitem{KNG}
C.A.R. Herdeiro, Class. Quantum Grav. \textbf{20}, 4891 (2003).

\bibitem{BDGO}
D. Brecher, U.H. Danielsson, J.P. Gregory, and M.E. Olsson, J. High Energy Phys. \textbf{11}, 033 (2003).

\bibitem{BC}
G. Barnich and G. Compere, Phys. Rev. Lett. \textbf{95}, 031302 (2005).

\bibitem{KV}
D. Klemm and L. Vanzo, Fortsch. Phys. \textbf{53}, 919 (2005).

\bibitem{KM}
R. Kerner and R.B. Mann, Phys. Rev. D \textbf{75}, 084022 (2007).

\bibitem{KA}
R.A. Konoplya and E. Abdalla, Phys. Rev. D \textbf{71}, 084015 (2005).

\bibitem{CGLP}
M. Cveti\v{c}, G.W. Gibbons, H. L\"{u}, and C.N. Pope,
hep-th/0504080.

\bibitem{BGHV}
E.K. Boyda, S. Ganguli, P. Horava, and U. Varadarajan, Phys. Rev. D \textbf{67}, 106003 (2003).

\bibitem{TO}
T. Ortin, Class. Quantum Grav. \textbf{22}, 939 (2005).

\bibitem{KLa}
H.K. Kunduri and J. Lucietti, J. High Energy Phys. \textbf{09}, 014 (2005).

\bibitem{BMPV}
J.C. Breckenridge, R.C. Myers, A.W. Peet, and C. Vafa, Phys. Lett. B \textbf{391}, 93 (1997).

\bibitem{GMT}
J.P. Gauntlett, R.C. Myers, and P.K. Townsend, Class. Quantum Grav. \textbf{16}, 1 (1999).

\bibitem{KS}
D. Klemm and W.A. Sabra, Phys. Lett. B \textbf{503}, 147 (2001).

\bibitem{CLP}
M. Cveti\v{c}, H. L\"{u}, and C.N. Pope, Phys. Lett. B \textbf{598}, 273 (2004).

\bibitem{MR}
O. Madden and S.F. Ross, Class. Quantum Grav. \textbf{22}, 515 (2005).

\bibitem{KLb}
H.K. Kunduri and J. Lucietti, Nucl. Phys. \textbf{B724}, 343 (2005).

\bibitem{Ross}
S.F. Ross, J. High Energy Phys. \textbf{01}, 130 (2006).

\bibitem{CCLP}
Z.W. Chong, M. Cveti\v{c}, H. L\"{u}, and C.N. Pope, Phys. Rev. Lett. \textbf{95}, 161301 (2005).

\bibitem{BTZbhT}
S. Wang, S.Q. Wu, F. Xie, and L. Dan, Chin. Phys. Lett. \textbf{23}, 1096 (2006).

\end{thebibliography}
\end{document}